# *Structural, thermal and optical tolerance analysis of SHARP, a near infrared spectrograph for next generation telescope*

**H. MAHMOODZADEH[1,2*], P. SARACCO[1], I. DI ANTONIO[3], P. CONCONI[1], D. SCACCABAROZZI[2], B. SAGGIN[7], E. MOLINARI[1], C. ARCIDIACONO[4], I. AROSIO[1], E. CASCONE[5], V. CIANNIELLO[5], V. DE CAPRIO[5], G. DI RICO[3], B. DI FRANCESCO[3], C. EREDIA[5], P. FRANZETTI[6], M. FUMANA[6], E. PORTALURI[3]**

[1]*INAF - Osservatorio Astronomico di Brera, Via Brera 28, Milano, Italy, 20121*
[2]*Politecnico di Milano, Department of Mechanical Engineering, Via la Masa 1, Milano, Italy, 20156*
[3]*INAF - Osservatorio Astronomico d'Abruzzo,Via Mentore Maggini, Teramo, Italy, 64100*
[4]*INAF - Osservatorio Astronomico di Padova, Vicolo dell'Osservatorio 5, Padova, Italy, 35122*
[5]*INAF - Osservatorio Astronomico di Capodimonte, Salita Moiariello 16, Napoli, Italy, 80131*
[6]*INAF - IASF, Via Alfonso Corti 12, Milano, Italy, 20133*
[7]*Università degli Studi di Padova Department of Industrial Engineering Padova Italy*
**Hossein.Mahmoodzadeh@Polimi.it*

**The design and verification of cryogenic near-infrared spectrographs require rigorous multidisciplinary analyses to ensure optical performance under realistic operational conditions. In this work, we present a structural–thermal–optical performance (STOP) analysis of the SHARP instrument, a multi-mode spectrograph designed for use with upcoming multi-conjugate adaptive optics (MCAO) systems on Extremely Large Telescopes (ELTs). Transient and steady-state thermal simulations were performed using Ansys to model the cooldown process and temperature distribution within the opto-mechanical structure. The resulting temperature fields were mapped to structural models through linear interpolation, enabling the evaluation of thermal stresses and deformations of optical elements. Optical sensitivity and tolerance analyses were carried out using Ansys Zemax OpticStudio, with Monte Carlo simulations applied to assess robustness against decenter, tilt, and despace errors. Results indicate that SHARP maintains RMS spot radius values within acceptance threshold, and that the inclusion of compensators significantly enhances compliance with the Ensquared Energy (EE) requirement across all channels. The analysis confirms that SHARP can achieves its optical performance goals while remaining within mechanical and thermal constraints, providing a validated framework for the development of next-generation cryogenic astronomical instrumentation.**

## 1. INTRODUCTION

SHARP is a multi-mode near-infrared (IR) spectrograph designed to exploit the capabilities of upcoming multi-conjugate adaptive optics (MCAO) systems on extremely large telescopes (ELTs) and other high-angular-resolution facilities. SHARP is composed of two primary units: NEXUS, a multi-object spectrograph (MOS), and VESPER, a multi-integral field unit (mIFU). Figure 1 illustrates the key components along the optical path from the port of MORFEO [1,2] to SHARP and the optical path of the light within SHARP. Briefly, the light from MORFEO enters SHARP after passing through the ADC. The ADC plays a crucial role by correcting for atmospheric dispersion, ensuring that SHARP can simultaneously collect spectra across the entire wavelength range, regardless of the zenith distance or rotation angle. Subsequently, the beam is directed by the unit selector system toward either the NEXUS or VESPER channel.

NEXUS is fed by configurable selecting system (CSS), which deploy up to 30 slits, each measuring approximately 2.4". The light is then folded by three folding mirrors, a design choice that compacts the instrument while maintaining optical precision. The light coming out of the CSS is split into four distinct wavelength bands, using three dichroic filters (Figure 1). Each band is fed by a dedicated camera hosting a grism wheel with three grism filter and an open position, with each camera optimized for its specific wavelength range. Four 2k×2k detectors are considered for each camera.

VESPER, the second main subsystem of SHARP, is an advanced multi-Integral Field Unit (mIFU) designed to perform spatially-resolved spectroscopy over the wavelength range of 1.2 to 2.4 μm. VESPER is specifically engineered to capture spectral information from multiple regions within a single field of view. The light coming from MORFEO is deviated toward the field selector system by the selector unit. Then, light enters VESPER via a series of folding mirrors and subsequently goes through optical elements before reaching the image slicer placed onto the VESPER focal plane (see Figure 1). The slicer placed on the focal plane of VESPER is composed of four sets of 72 micro-mirrors each. Each set samples a stripe equal to 1/4 of the image and sends the light to a camera summing up to four cameras per module. VESPER is a modular system. Each module is composed of 6 probes called Field Selectors (FSS). The current configuration of VESPER includes 2 modules summing to 12 FSs and to 8 cameras, each fed by a 4kx4k detector.

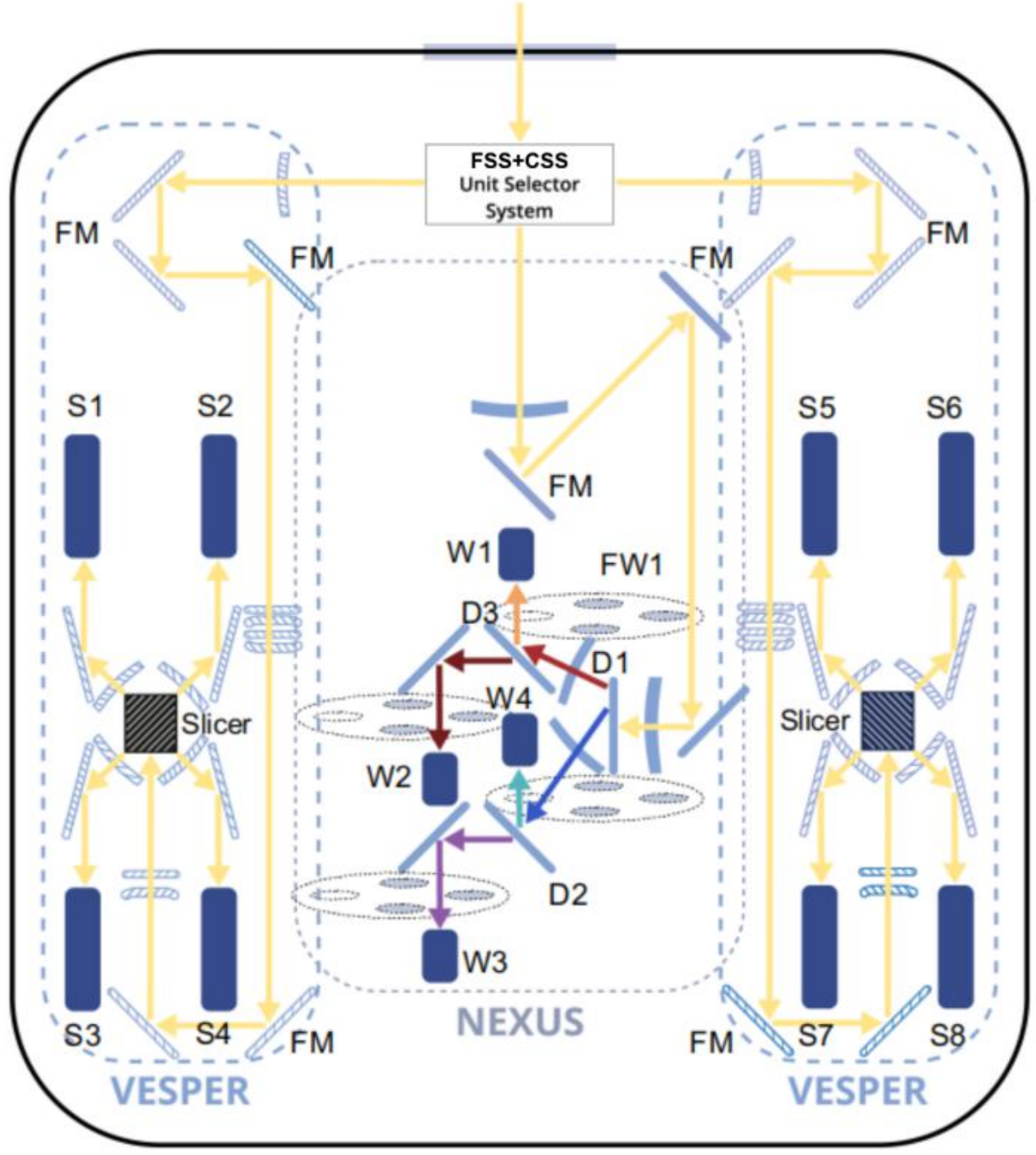

**Fig. 1.** Components and optical path from MORFEO to SHARP and optical path of SHARP and its key components. SHARP is fed by an ADC and an NGS unit.

The scientific objectives and optical–mechanical design of SHARP have been detailed in previous works by Saracco et al. (2024) [3], and Mahmoodzadeh et al. (2025) [4]. The scientific requirements for SHARP are summarised in Table 1. SHARP represents a paradigm shift as the first multi-object and multi-IFU spectrograph designed to exploit wide-field adaptive optics on the ELT. Demonstrating the feasibility of this concept poses a unique engineering challenge due to the unprecedented complexity and sheer size of its components. Crucially, despite the massive scale of the instrument and of its components, its mechanical moving parts and overall structure must maintain extreme positional accuracy to fully leverage the diffraction-limited capabilities provided by the ELT's AO systems. Therefore, the primary novelty of this work lies in our integrated modeling approach. Evaluating optical tolerances, mechanical flexures, thermal stress, and the implementation of compensators under these extreme boundary conditions requires a pioneering workflow. The analysis strategy presented in this paper goes beyond standard instrument design, providing a novel framework by introducing a computationally efficient STOP methodology based on appropriate assumptions and simplifications. This approach significantly reduces the computational cost of the analysis while enabling reliable validation of the structural and optical feasibility of next-generation, AO-assisted giant instruments. Both space and ground-based optical systems routinely employ coupled structural–thermal–optical performance (STOP) analyses as part of integrated multidisciplinary modelling frameworks. STOP methods have been applied to large observatory subsystems and instruments to quantify thermo-elastic deformation, predict optical performance under operational loads, and guide the specification of tolerances and compensators. For example, Holzlöhner et al. performed STOP studies on subsystems of the European Extremely Large Telescope [ELT] [5], and Chengliang et al, applied similar techniques to a cryogenic infrared spectrometer [6]. Finite-element modeling has been used to investigate the structural and thermal behavior of EMIR [7], while a combined STOP and experimental verification was reported for the HYPSO hyperspectral imager on a CubeSat platform [8]. Major space projects such as the James Webb Space Telescope (JWST) and its integrated instrument modules [9,10], Herschel telescope [11] and Giant Magellan Telescope [12,13] have relied on thermal and thermo-elastic analyses. In addition to thermal analysis, other structural analyses are implemented during design and verification: Manil and Lieber (1997) report performance modelling for the ESO Very Large Telescope (VLT) that includes wind, gravity and seismic loading [14]. Collectively, these efforts demonstrate that integrated STOP workflows are essential for ensuring optical systems meet demanding performance requirements in realistic operational environments.

**Table 1: SHARP main requirements resulting from the scientific drivers.**

| | |
|---|---|
| Wavelength limit | 2.45 µm |
| Spectral ranges (simultaneous) | 0.95-2.45 µm |
| Angular resolution | ~ 30 mas |
| AO-corrected FoV | ~ 1x1 arcmin |
| Observing modes | MOS + multi-IFU |
| Spectral resolution | R>1000 |

The workflow of analysis began with optical sensitivity and tolerance analyses performed in Zemax Optic Studio. through Monte-Carlo simulations, these analyses were used to determine the maximum allowable rigid-body perturbations of the optical elements, including decenter, tilt, and despace errors, while maintaining the required optical performance. The obtained limits defined the optical error budget of the system and were later used as acceptance criteria for the thermo-mechanical analysis and to define compensation and alignment strategies. Thermal simulations were then carried out in ANSYS to evaluate both the transient cooldown process and the steady-state operational condition. The transient analysis was used to estimate the required cooling power, cooldown duration, thermal gradients, and the effectiveness of the thermal bridges and cooling interfaces. The steady-state analysis was performed to determine the final operational temperature distribution within the structure and optical subsystems. The temperature distributions obtained from the thermal simulations were subsequently applied to the structural finite-element model together with gravity loading conditions. Structural analysis was then performed to calculate the rigid-body displacements and rotations of the optical components under operational cryogenic conditions. Since the optical elements are mounted using compliant interfaces intended to minimize thermo-mechanical stress transfer from the structure to the optics, the analysis focused primarily on rigid-body motions rather than detailed surface deformation of the optical components. In the adopted STOP methodology, the optical elements were represented in the structural model using concentrated point-mass elements positioned at their respective centers of mass. This approach was selected to reduce computational cost while still accounting for the gravitational contribution of the optical elements to the supporting structure. The resulting structural deformations were directly compared with the allowable perturbation limits derived from the Zemax tolerance analysis to verify compliance with the optical error budget.

## 2. TOLERANCE ANALYSIS

The optical paths of SHARP for the NEXUS and VESPER modes are shown in Figures 2 and 3, respectively. Further details regarding the optical paths and lens specifications can be found in Mahmoodzadeh et al (2025) [4]. The tolerance analysis for both NEXUS and VESPER uses the RMS spot radius relative to the centroid as the primary metric, closely aligned with the system's 80% Ensquared Energy (EE) preliminary requirement (the use of the EE metric is not convenient for analyses such as tolerance sensitivity and system optimization, since its computation is slower and less precise than that of the RMS spot radius).

### A. NEXUS Tolerance Analysis

The upper limit for the rms spot radius is set at 7.5 µm, corresponding to half the detector pixel size. The analysis considers nine fields within the 72”×72” FoV: (0,0), (±16″, ±16″), (0, ±36″), and (±36″, 0), each weighed equally, with corner fields excluded. Initially, alignment tolerances were evaluated on the common-path optics down to the first dichroic lens (DICR.1 as shown in Figure 2). Given the limited margin with respect to the maximum allowable spot size and the significant number of optical surfaces, tight mechanical tolerances were applied: lens decenter ±0.1 mm, lens tilt ±0.01°, mirror decenter ±0.5 mm, mirror tilt ±0.01°, and despace ±0.2 mm. Monte Carlo simulations (500 runs, 9 fields × 3 wavelengths, >90% statistical coverage) showed that the system is generally insensitive to decenter and tilt of the window (is not shown) and field lens (L1), with an average spot radius increase of ~0.2 µm across most channels, except for a higher sensitivity noted in the J channel.

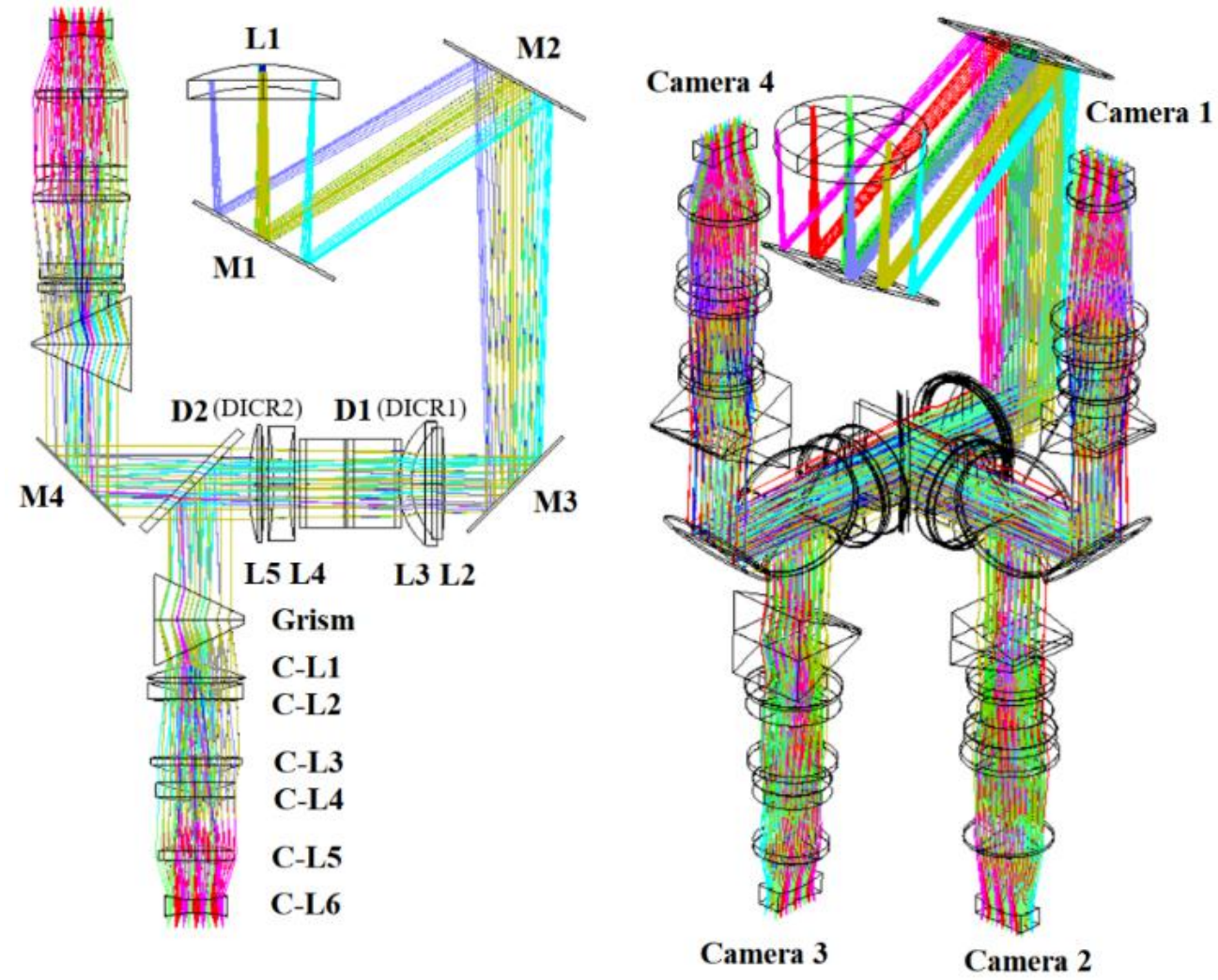

**Fig 2** The optical path of NEXUS, from the focal plane down to the individual cameras (the window is not shown). (left) section view of the optical path (right) 3D view of the optical path. In the left for simplicity only two channels are shown (the other two channels including DICR3 and M5 and its relevant cameras is not shown in 2D).

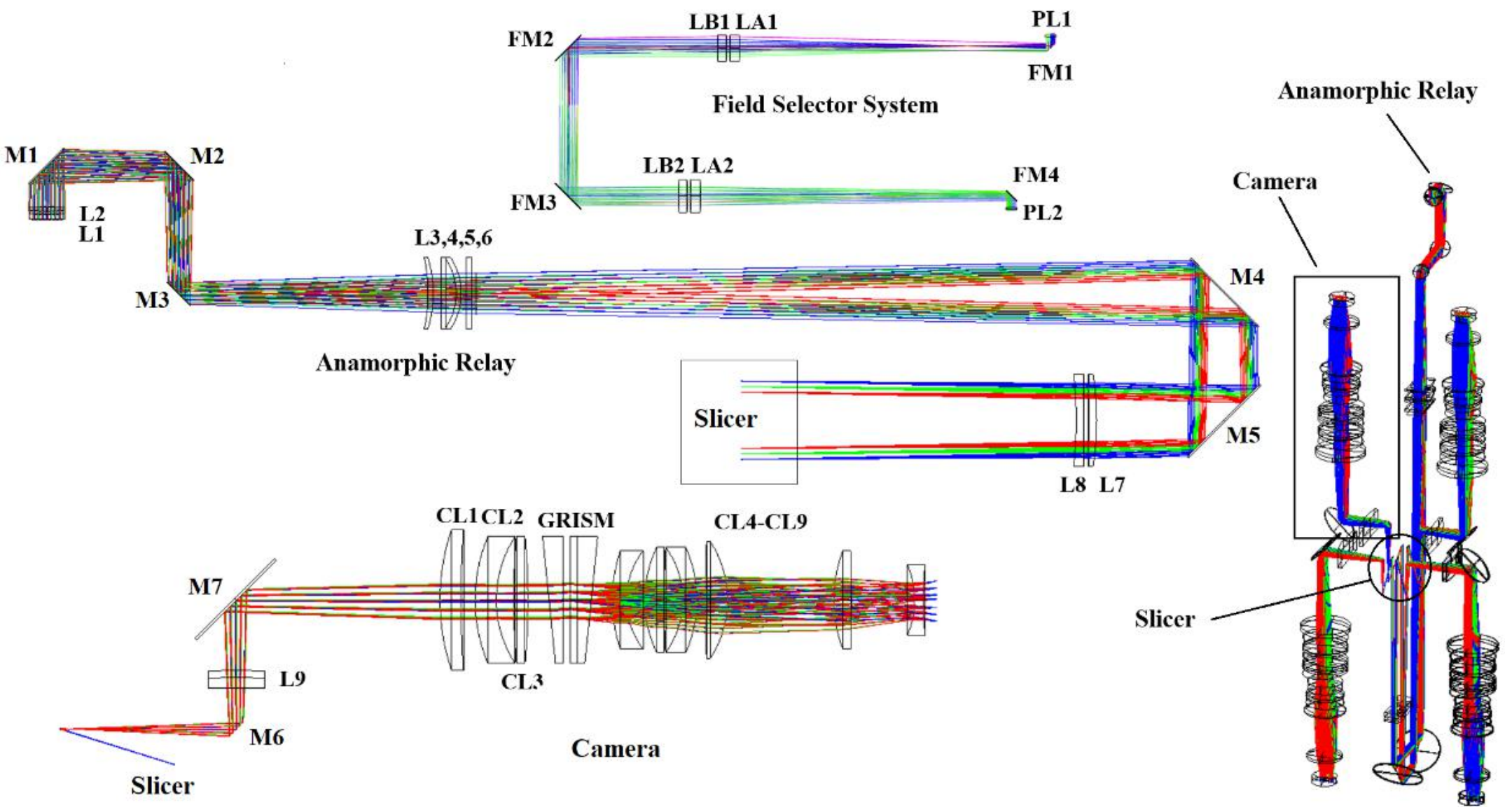

**Fig 3** Top: the Field Selector System (FSS), Mid: the Anamorphic Relay and the Slicer, Bottom: the Camera optics

In the second stage, the analysis included the K-H and J-Y paths (L4, L5, DICR2 and L7, L8, DICR3) while retaining the initial tolerance values. The H and K channels exhibited minimal additional sensitivity, while the J and Y channels showed slight increases (~0.2 µm) due to L4, L5, DICR.2 and L5, L6, DICR.3 tilt and decenter contributions, identifying new critical elements requiring monitoring during integration.

The third stage incorporated the GRISM and camera optics, which dominate the sensitivity landscape. Each camera is planned as a pre-aligned, stand-alone unit, allowing tight tolerances: lens decenter ±0.02 mm (±0.01 mm for CL3, CL4), lens tilt ±0.005° (±0.003° for CL3, CL4), and despace ±0.02 mm. The H and J channels demonstrated high sensitivity to camera lens tolerances, requiring further tightening for CL3 and CL4 to achieve near-spec performance, while the K and Y channels remained within manageable limits under the defined tolerances.

To further reduce sensitivity and maintain compliance with the EE requirements, effective compensators were evaluated. Besides the detectors' position (focus adjustment), the tip/tilt of M2 (general alignment), DICR.2,3 (for H and J), and M4,5 (for K and Y) were introduced as compensators in the simulations. Final Monte Carlo results demonstrated reduced spot sizes across channels: 5.4 µm (K), 6.6 µm (H), 7.2 µm (J), and 5.8 µm (Y). These results confirm that K requires only focus compensation, while H and J benefit from DICR.2,3 and M2 adjustments, and Y leverages FM and M2 for optimal alignment. The flow chart of the analysis process is shown in Figure 4.

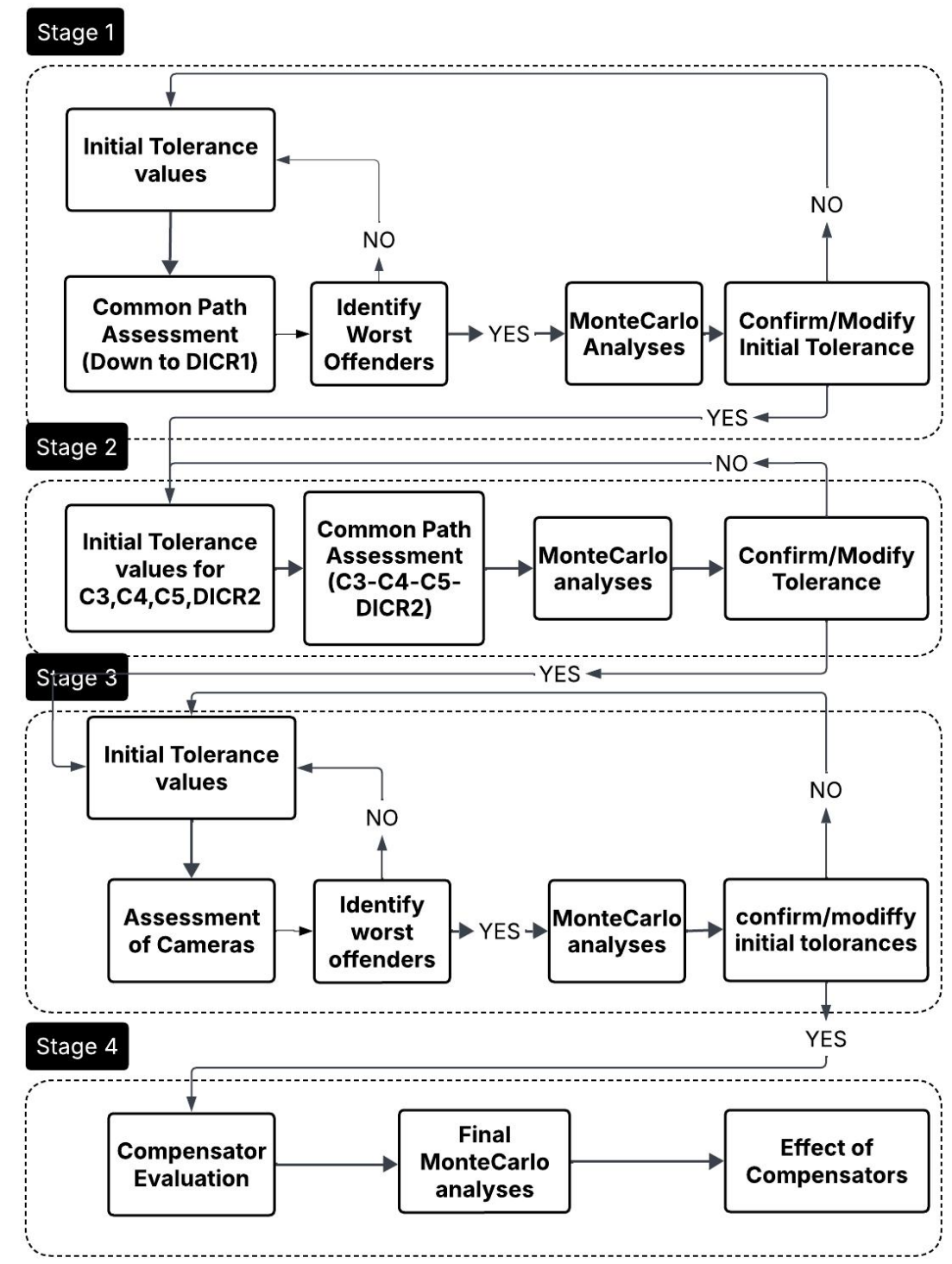

**Fig 4.** NEXUS Tolerance Analysis Workflow. Schematic illustrating the structured tolerance analysis approach for NEXUS, including common-path assessment, camera optics integration, Monte Carlo simulation, and compensator evaluation for cross-channel compliance

The structured approach ensures that NEXUS maintains optical performance within specifications while systematically managing tolerance and alignment constraints across all channels.

The results of the tolerance analysis are summarised in Tables 2–4. Table 2 reports the mechanical tolerance ranges applied during the alignment analysis. The outcomes of the Monte Carlo simulations without compensators are shown in Table 3. Table 4 presents the results after applying effective compensators, including the tilts of M2, DICR2, M4,5, besides the focus adjustment. These compensators significantly reduce the RMS spot radius across all channels, enabling the system to recover optical performance.

The Monte Carlo simulation outcomes are further illustrated in Figures 5 and 6. Figure 5 presents the RMS spot radius distributions for the H channel, comparing cases with and without compensators. The results indicate that most realizations already fall within the 7.5 µm acceptance threshold, and compensators provide only a marginal improvement for this channel. In contrast, Figure 6 demonstrates the significant impact of compensators on the J channel, where the distributions shift toward smaller spot radii, substantially enhancing compliance with the ensquared energy requirement.

**Table 2: Tolerance Ranges Applied During Alignment Analysis. Summary of mechanical tolerances applied in the NEXUS tolerance analysis, including lens decenter, lens tilt, mirror decenter, mirror tilt, and despace values used for Monte Carlo simulations.**

| Parameter | Value (±) | Notes |
|---|---|---|
| Lens decenter | 0.1 mm | Applies to all lenses |
| Lens tilt | 0.01° | Applies to all lenses |
| Mirror decenter | 0.5 mm | Flat mirrors only, no sensitivity |
| Mirror tilt | 0.01° | Applies to all mirrors |
| Lens/Mirror despace | 0.2 mm | Inter-element distance tolerance |
| Added Grism + Camera optics | | |
| Grism Rotation | 0.5° | clocking angle |
| Lens decenter | 0.02 mm | 0.01 mm for CL3 and CL4 |
| Lens tilt | 0.005° | 0.003° for CL3 and CL4 |
| Lens despace | 0.02 mm` | Inter-element distance tolerance |

**Table 3: Monte Carlo Simulation Results Without Compensators. Average RMS spot radius (µm) for K, H, J, and Y channels after 500 Monte Carlo runs (>90% statistical coverage) under defined tolerance conditions without additional compensators, indicating sensitivity levels per channel.**

| Channel | MF In (µm) | MF Fin (µm) | Sensitivity Observations | Compliance |
|---|---|---|---|---|
| K | 5.0 | 5.4 | Minor increase, stable | Pass |
| H | 5.8 | 6.9 | Higher sensitivity to tilt and decenter (L2, L1) | Marginal |
| J | 6.1 | 8.7 | High sensitivity (tilt L3, M1, M2 and decenter L3, L4) | Fail |
| Y | 4.7 | 6.7 | Moderate sensitivity, primarily tilt and decenter L4 | Pass |

**Table 4: Monte Carlo Simulation Results with Compensators Applied. Average RMS spot radius (µm) for K, H, J, and Y channels after implementing effective compensators (focus mechanism, M2, DICR.2, FM) alongside the defined tolerance sets, demonstrating reduced spot sizes across channels.**

| Channel | MF In (µm) | MF Fin Before (µm) | MF Fin After (µm) | Main Compensators Applied | Compliance |
|---|---|---|---|---|---|
| K | 5.0 | 5.4 | 5.3 | Focus | Pass |
| H | 5.8 | 6.9 | 6.6 | Focus, M2, DICR2 | Pass |
| J | 6.1 | 8.7 | 7.2 | Focus, M2, DICR2 | Near Pass |
| Y | 4.7 | 6.7 | 5.8 | Focus, M2, M4,5 | Pass |

## B. VESPER Tolerance Analysis

The general strategy for the VESPER is similar to the NEXUS, but in this case, we have eight identical paths, so one single model is set and analysis repeated for two of the 72 system configurations (one for each slicer) and for two wavelengths within the range 1.4-2.2 µm. As for NEXUS the upper limit is set at 7.5 µm, corresponding to half the detector pixel size. The analysis considers three fields defined for each configuration, the central weighted double.

In the first step, the optical path from the entrance window up to the FSS output, comprising the optical elements W, PL1, LA1, LB1, LB2, LA2, and PL2, was investigated. Despace (inter-distance) tolerances of 0.04 mm were applied for short separations (LA1–LB1, LB2–LA2) and 0.08 mm for longer distances. Lenses decenter tolerances of 0.03 mm were applied for LB1 and LB2, and 0.05 mm for the remaining lenses, with a lens tilt tolerance of 0.01°. A total of 500 Monte Carlo runs were performed using parabolic statistics, with the merit function defined as the spot size averaged over three fields, and a focus compensator applied to maintain optimal image quality. The results show that the system is not highly sensitive to the tip/tilt of the FSS optics, suggesting that lens-tilt tolerances could be relaxed. However, the spot radius increases at longer wavelengths (where the nominal performance is worse) is more significant, up to a factor of 2.

The second step extends the analysis by adding the Anamorphic Relay, including optical elements L1, L2, M1, M2, M3, L3–L8, and M4, M5.

.

**Fig. 5** Monte Carlo spot radius distributions for the H channel. The left panel shows results without compensators applied, while the right panel shows results with compensators. The vertical dashed line indicates the 7.5 µm acceptance threshold, illustrating the limited improvement when compensators are applied.

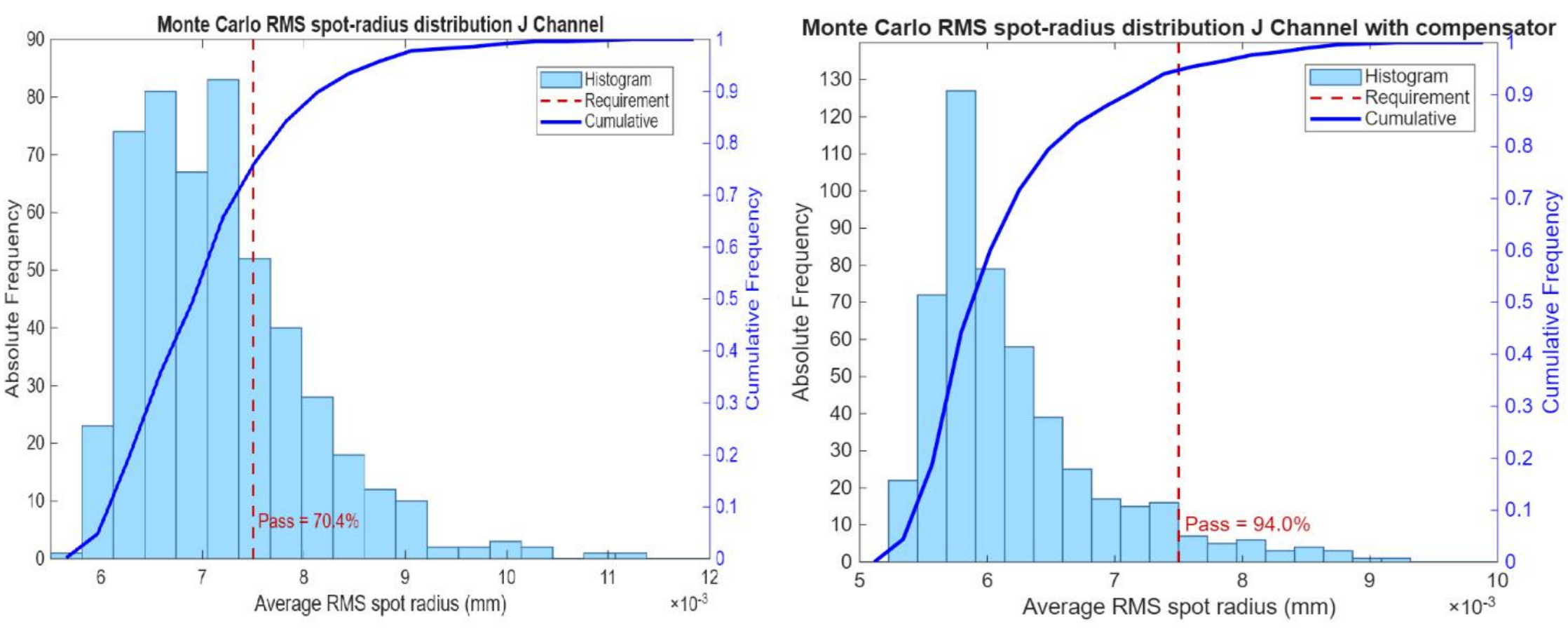

**Fig. 6** Monte Carlo spot radius distributions for the J channel. The left panel shows results without compensators applied, while the right panel shows results with compensators. The vertical dashed line indicates the 7.5 µm acceptance threshold, illustrating a significant improvement when compensators are applied.

The applied tolerances are as follows: despace = 0.1 mm between the FSS and Anamorphic Relay, 0.05 mm between lenses, and 0.2 mm for all others; lens decenter = 0.03 mm for L2 and 0.1 mm for the rest; lens tilt = 0.01°; mirror tilt =0.005°; and biconic lenses (Lx) rotation ≤ 0.1°. Results indicate that the system is quite sensitive to the tip/tilt of the flat mirrors, necessitating compensators while being relatively insensitive to lens tilt.

Step 3 extends Steps 1 and 2 by incorporating the slicer and pupil mirrors. Since adjustments or compensators cannot be applied to these components, all slicer errors are treated as fixed residuals. Analysis demonstrates that the system is generally insensitive to slicer tolerances, particularly at longer wavelengths, if provided tolerances are sufficiently tight. Namely: despace = 0.1 mm, mirror decenter = 0.04 mm, mirror tilt = 0.01° for pupil mirrors, and mirror tilt = 0.005° for slicer mirrors.

Step 4 considers the full optical system, including all the remaining optical elements (M6, L9, M7, CL1–CL3, GRISM, and CL4–CL9). The tolerances for the pre-camera and camera optics are consistent with those defined for NEXUS: lens decenter = 0.02 mm (0.01 mm for CL4, CL5, and CL7), lens tilt = 0.005°, lens despace = 0.02 mm, and mirror tilt = 0.005°. The analysis indicates that the system is sensitive to tilt and decenter of some camera optics (CL4, CL5, CL7), especially at shorter wavelengths.

In conclusion, the overall sensitivity of the system also depends on the wavelength. At longer wavelengths, the system exhibits higher sensitivity to the tolerances defined in the first part (FSS and Anamorphic Relay), whereas at shorter wavelengths the sensitivity to the tolerances of both the first (FSS + Anamorphic Relay) and second parts (Slicer + Camera) are lower and more balanced. The tolerances of the Anamorphic Relay mirrors, if uncompensated, have a strong influence on the system performance, particularly at longer wavelengths. To ensure a proper beam alignment toward the slicer and to minimise the overall sensitivity, at least one (preferably two) mirrors in the Anamorphic Relay (specifically M3 or M5) shall include tip/tilt adjustment mechanisms, the summary of the tolerance analysis results for the analysed configurations (config.11 and config.42) is presented in Table 5. The effect of the compensators mirrors is presented in
Table 6.

**Table 5: Comparative summary of Monte Carlo tolerance analysis results for 2 of the 72 configurations of VESPER. The table presents RMS spot-radius performance and system sensitivity across all optical subsystems (FSS, Anamorphic Relay, Slicer, and Camera), highlighting configuration-dependent variations in tolerance sensitivity and overall optical performance.**

| | CONF.42 | CONF.11 | CONF.42 | CONF.11 |
|---|---|---|---|---|
| **Wave length** | **1.6 um** | **1.73 um** | **2 um** | **2.2 um** |
| **MF in (um)** | **3.8** | **3.2** | **6** | **5.6** |
| | | **Stage 1** | | |
| MF fin (um) | 4.4 | 3.9 | 7.3 | 6.9 |
| | | **Stage 2** | | |
| MF fin (um) | 5.5 | 4.7 | 7.7 | 7.8 |
| | | **Stage 3** | | |
| MF fin (um) | 5.7 | 4.8 | 8.4 | 7.7 |
| | | **Final analyses** | | |
| MF fin (um) | 6.6 | 6.1 | 9.1 | 8.4 |
| Delta MF (um) | +2.8 | +2.9 | +3.1 | +2.8 |
| Worst offenders | Dec: CL5, CL7 Tilt CL4, CL5, CL6, CL7 Rot: L3, L4 | Rot: L3, L4 Dec: L1, CL4, CL7 Desp L1, L2, PL2-L1, LA1-LB1, LB2-LA2 | Tilt: L4, CL5, M1, M3, M2 Dec: L1, CL7 Rot: PL2 | Tilt: M3, M2, M1 same as before |

**Table 6: Effect of compensator (tilt of mirror M5 of the Anamorphic Relay) on system performance, highlighting improvements in RMS spot-radius reduction compared to the uncompensated configuration.**

| | CONF. 42 | CONF. 11 | CONF. 42 | CONF. 11 |
|---|---|---|---|---|
| | **1.6 um** | **1.73 um** | **2 um** | **2.2 um** |
| Compensators | - | - | Tilt M5 | Tilt M5 |
| Delta MF (um) | - | - | +2.6 | +1.6 |

## 3. Finite Element Analysis

The thermal and structural analysis of the SHARP system was carried out using the commercial software package ANSYS 2023. The initial 3D geometry was created in SolidWorks and then imported into ANSYS for finite element modelling (FEM). AL 6061 T6 is considered for the internal structure, and stainless steel 304L for the external vacuum tank. SHARP is giant instrument with 2500 mm in diameter and 3300 mm in height so to reduce model complexity, enhance simulation accuracy, and minimise computation time, several simplifications were applied including: Minor features such as small holes and fillets were removed, Fastening elements including screws, nuts, and washers were excluded, Optical components were represented as point masses using remote nodes.

Figure 7 illustrates the simplified model of the SHARP system prepared for simulation.

The mesh was created using a combination of tetrahedral and hexahedral elements, with higher mesh density applied to the internal structure in critical areas, as regions experiencing high thermal gradients or mechanical stress, and a coarser mesh was applied to the storage tank as non-critical regions to reduce computational time while maintaining accuracy. A mesh-convergence assessment was performed by reducing the element size until the variation in the structural deformation results became negligible within the required accuracy range for the optical performance evaluation. The applied mesh is shown in Figure 8.

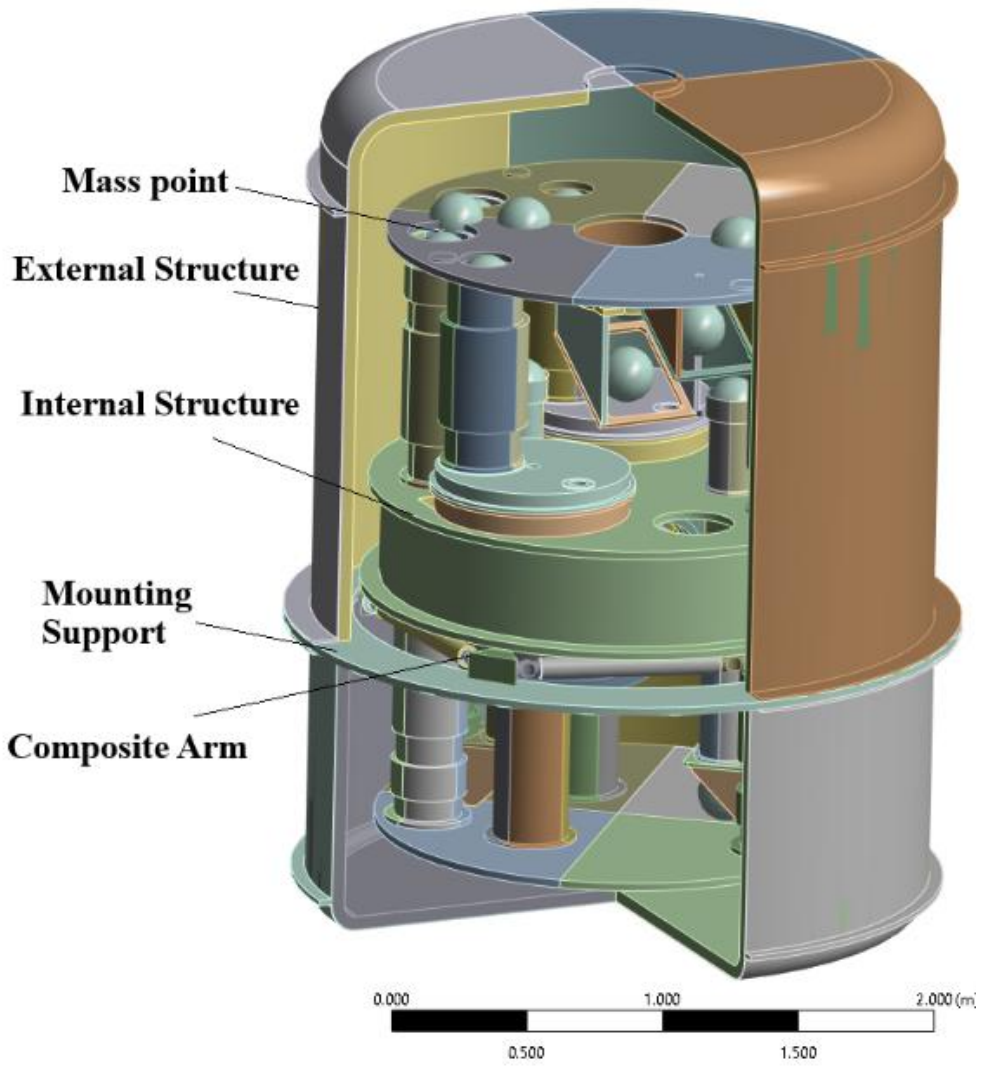

**Fig 7.** Simplified 3D model of the SHARP system used for thermal and structural analysis. Minor features were removed, and optical components were represented as point masses to optimize simulation efficiency and reduce computational load.

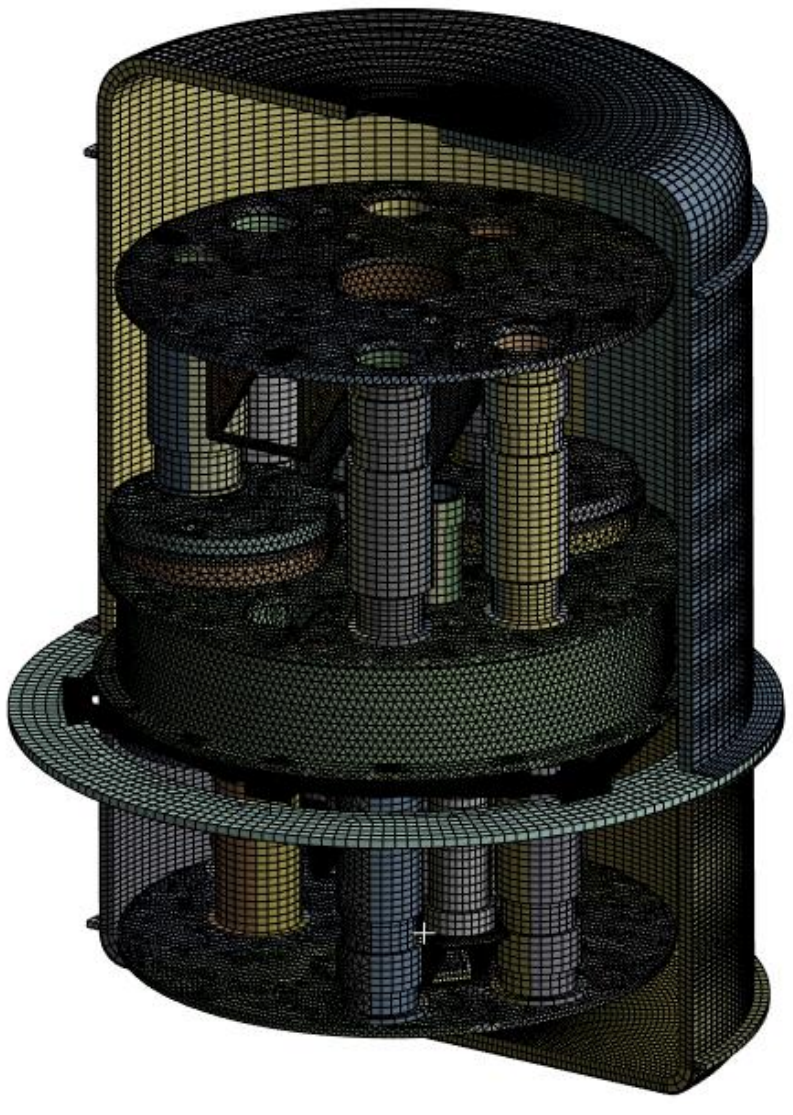

**Fig. 8** Finite element mesh of the SHARP system, a combination of tetrahedral and hexahedral 3D solid elements is used to prepare the FEM model. The model includes 663870 elements and 330544 nodes.

#### A. Thermal analysis

Based on the scientific requirements of SHARP [3], the system must operate at cryogenic temperatures to minimize thermal background noise. Specifically, the optical components should be maintained at maximum 80 K. To achieve these temperature targets, a total of 8 cryocooler heads are employed, four mounted on the upper surface and four on the lower surface of the internal structure. The arrangement aims to maintain consistent thermal conditions throughout the system, minimizing temperature gradients. The arrangement of the cryocooler heads is illustrated in Figure 9. The specification of the cooling probe is illustrated in Figure 10.

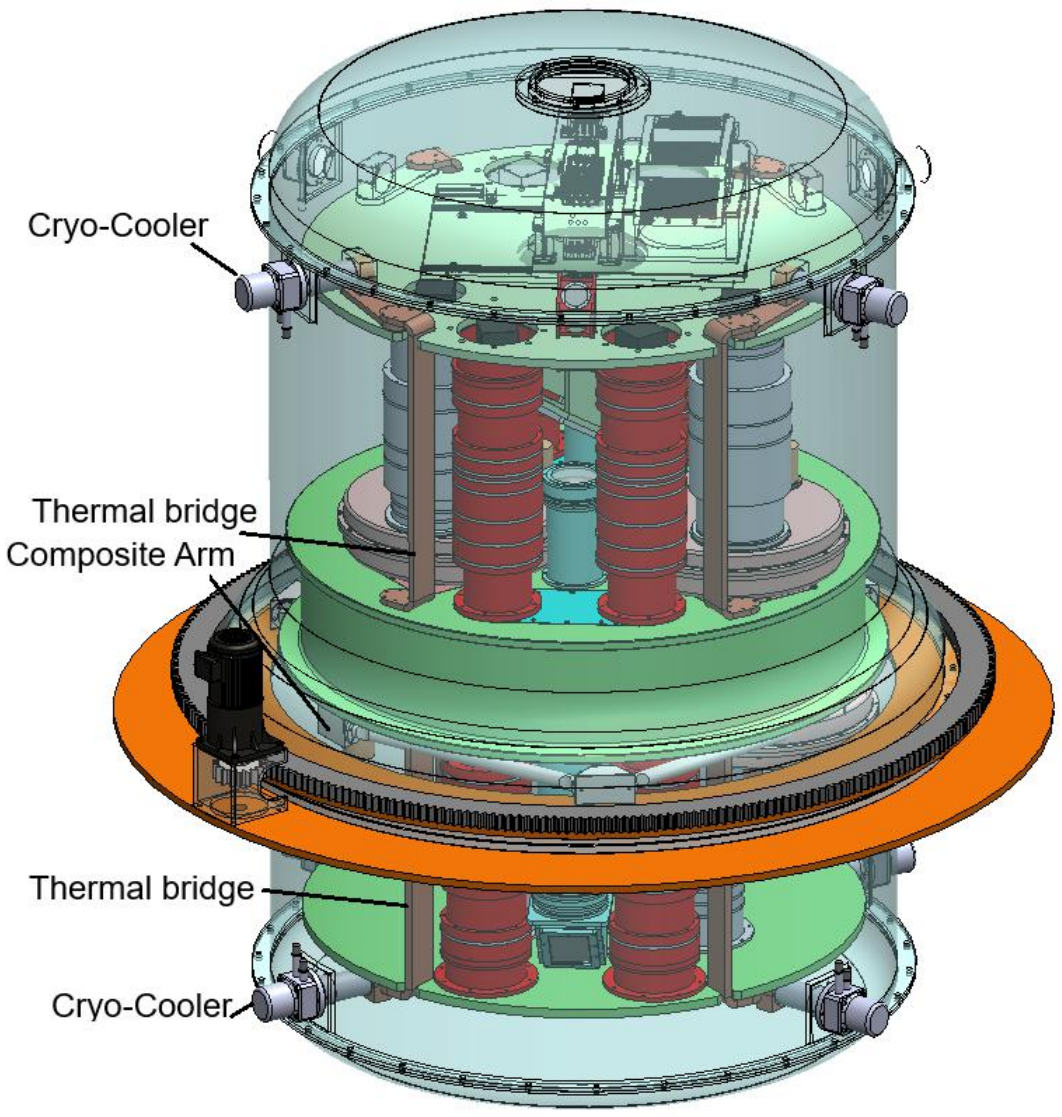

**Fig. 9:** Schematic view of the eight cryocooler heads mounted on the structure, with four located on the upper surface and four on the lower surface, ensuring uniform thermal distribution and minimized temperature gradients across the system.

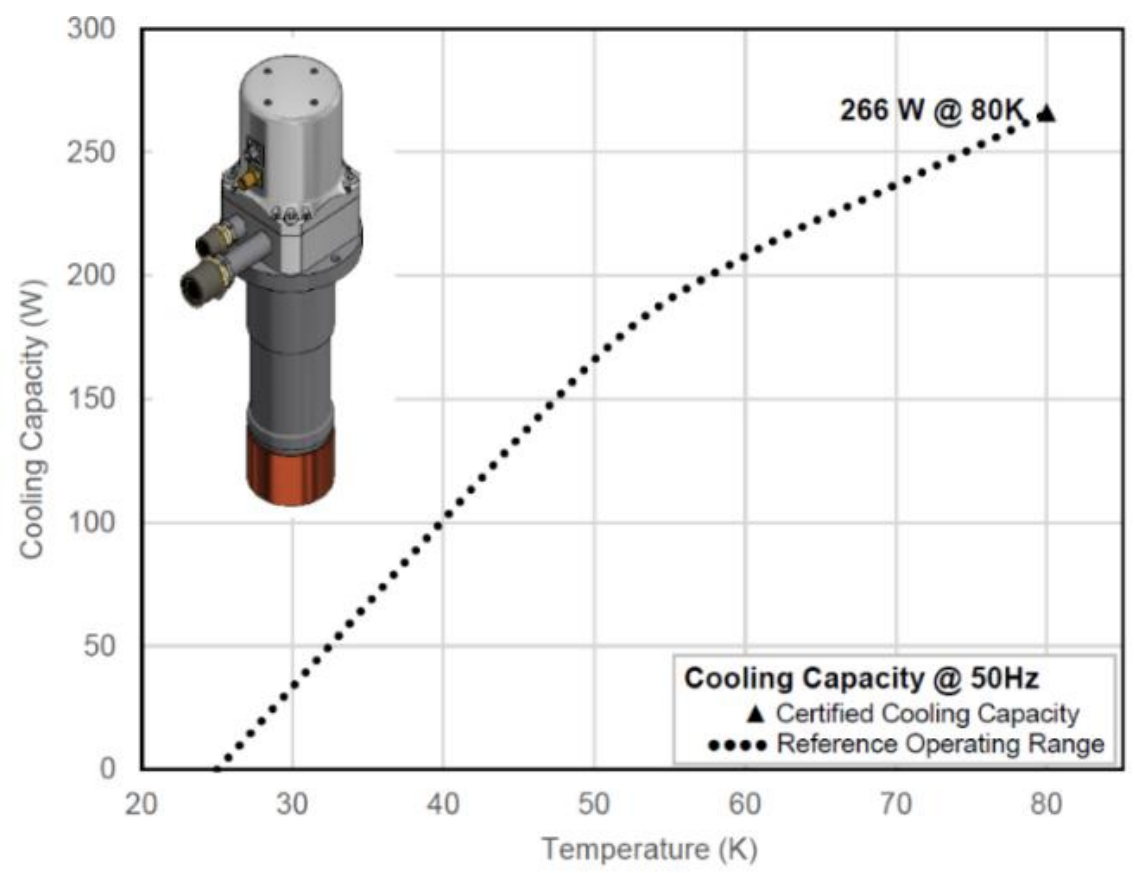

**Fig. 10:** Technical specifications of the cryogenic cooling probes employed in SHARP

*1.Transient analysis*

To investigate the cooling process of the structure and to estimate the time required to reach the desired temperature, and determine the required number and optimal locations of the thermal bridges, a transient thermal analysis was performed on the SHARP structure. The thermal input to the system is applied as a combination of heat flux and fixed temperature as boundary condition on designated surfaces. Power losses occur through three primary mechanisms: convection from the external tank, radiation, and conduction through the composite arm. During the transient analysis, unlike in the steady-state case, all internal electric engines and equipment are turned off, resulting in no internal heat generation.

Implementing radiation boundary conditions in ANSYS can significantly increase solver complexity and computational time. To simplify the simulation while maintaining acceptable accuracy, the radiation losses are approximated using an equivalent heat flux.

Based on established references in cryogenic engineering, the loss heat flux through a well-designed multilayer insulation (MLI) system under high-vacuum conditions typically ranges between 0.1 and 1.0 W/m² [15-17]. These values assume a vacuum pressure below $10^{-5}$ torr and the use of 20 to 30 reflective layers with low-emissivity materials (within the range of 0.01 and 0.02). In our case, the heat flux through the MLI insulator is approximated using a fixed value of 0.3W/m2 (i.e. accounting for the radiation and residual convective heat loads estimation).

In this simulation, a constant cooling flux (~2000 W/m2) is initially applied at the probe contact point until the temperature reaches 70 K. Then, the heat flux is replaced with a fixed temperature boundary of 70 K, allowing the structure to cool according to the heat propagation speed within the material. This approach maintains the contact point at 70 K, consistent with the real condition in which the control system keeps the contact point at a minimum temperature of 70 K based on the cooling system capacity. The radiation and heat losses due to the residual air after vacuum application are applied as a fixed heat flux on the external surface of the internal structure of the spectrograph.

In the initial configuration, heat was extracted from the structure through eight thermal contact points (four on the top surface and four on the bottom) without the inclusion of thermal bridges (see Figure 9). At this stage, it was assumed that the cooling system could deliver its maximum capacity, corresponding to ~250 W per probe and a total of 2000 W. Any reduction in available cooling power would accordingly increase the overall cooling time. The temperature distribution within the structure over time is illustrated in Figure 11, while the average temperature of the main structure (cylindrical structure on which all cameras are installed) as a function of time is presented in Figure 12. As shown, the system reaches a steady-state temperature after 125 hours, with a minimum temperature of 79 K.

The heat flux extracted by each cooling probe over time, along with the temperature at the contact point and the mean temperature in the main box, is shown in Figure 13. The maximum heat flux applied to the system is 245 W, which remains below the maximum cooling capacity of the cryogenic system. As evident from the figure, the temperature decreases to 70 K after approximately 34 hours. To maintain the structure at this temperature, the applied heat flux decreases significantly, and the system reaches steady-state conditions after about 125 hours.

To improve the cooling performance, eight thermal bridges were added as additional heat extraction points on the main unit; the thermal bridges are shown in Figure 9. In this configuration, as shown in Figure 14, the system reaches steady-state conditions after 68 hours, and the structure can uniformly reach 70 K. The corresponding heat flux curve is shown in Figure 15. It is worth noting that in this case as well, the heat-flux curve remains below the maximum cooling power. However, the cooling system operates at its maximum capacity for 50 hours (more than 90% of the total cooling time) in order to cool down the structure in a shorter period and in optimum condition.

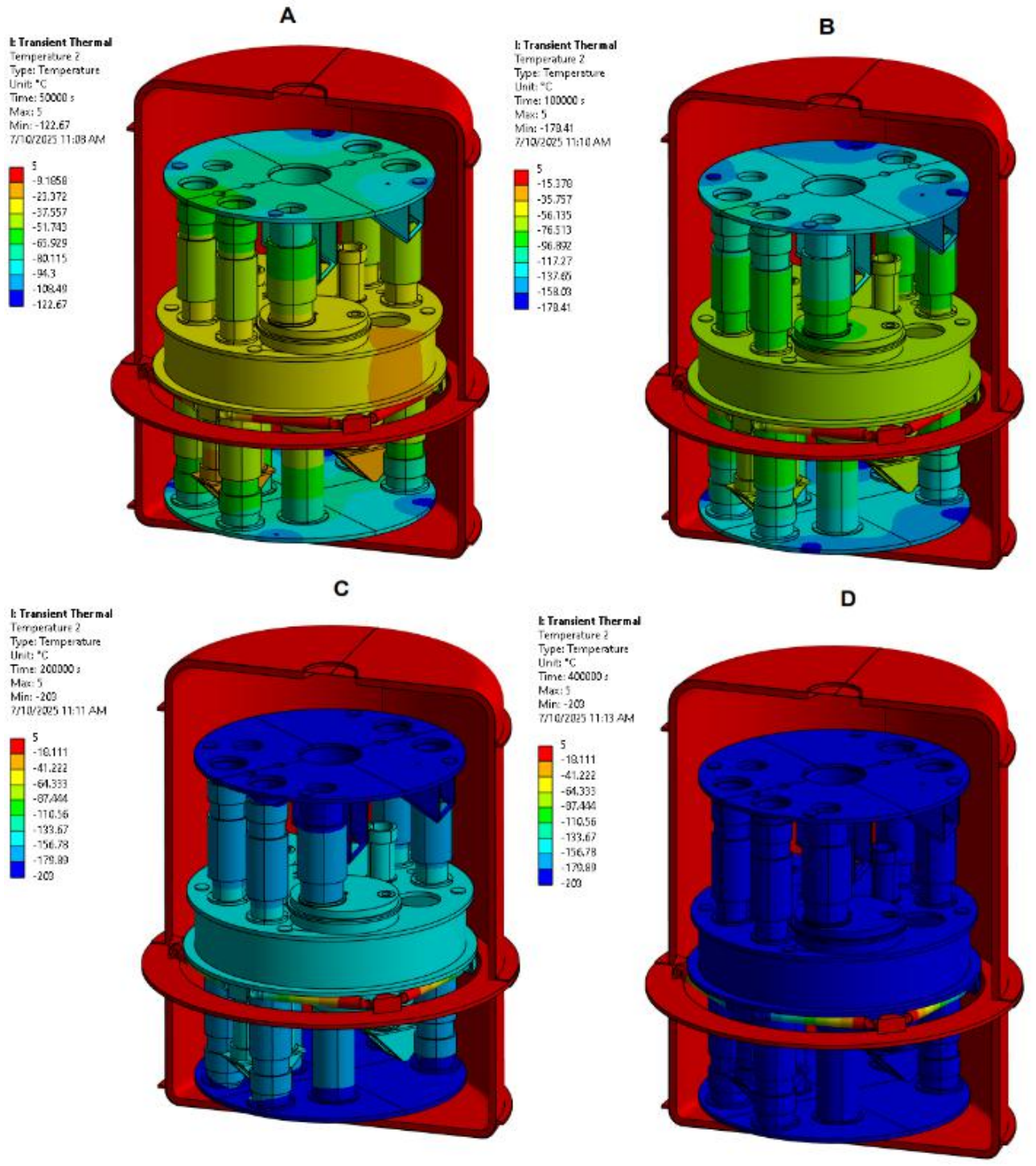

**Fig. 11** Temperature distribution within the structure over time during cooling using eight thermal contact points (four on the top surface and four on the bottom surface).

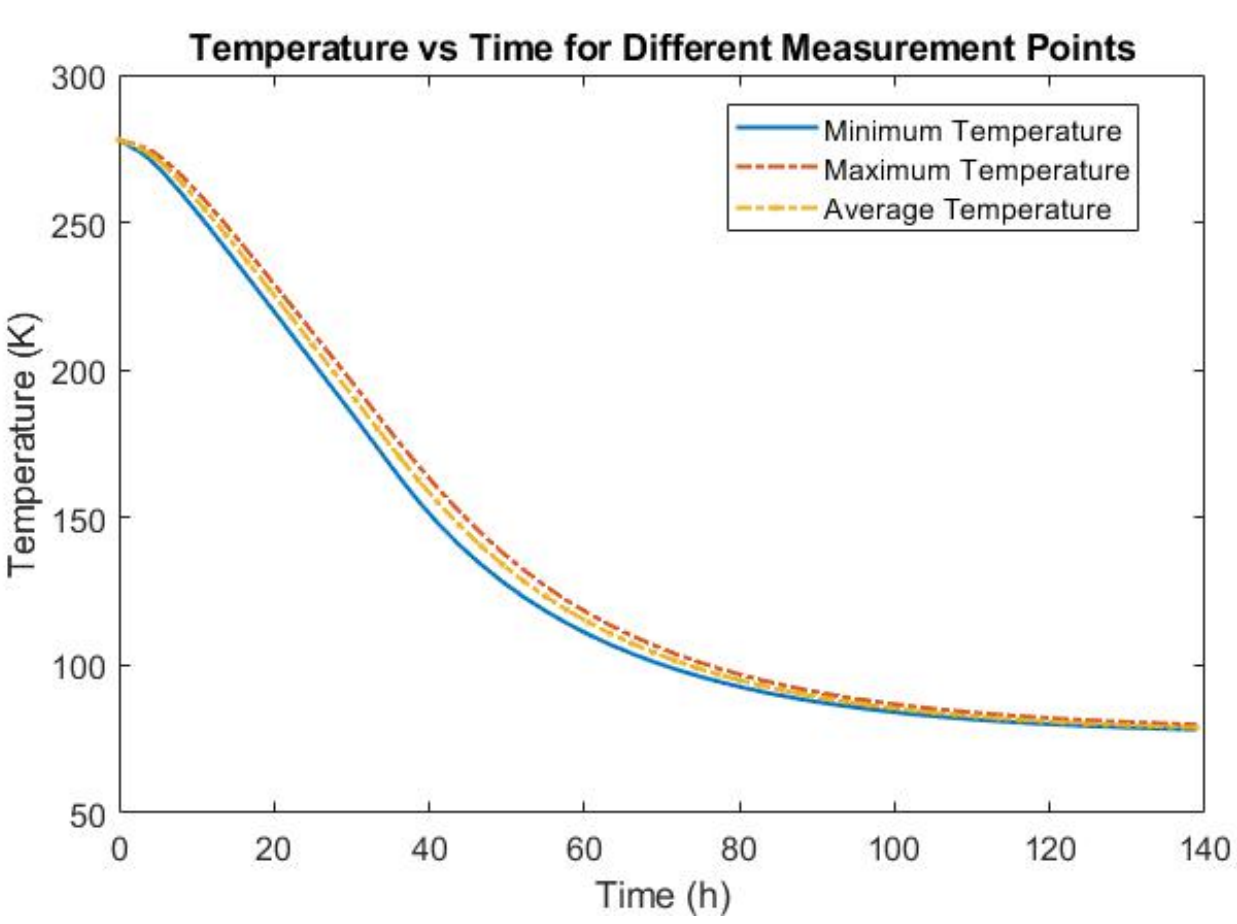

**Fig. 12** Average temperature of the main structure as a function of time during the cooling process. The system reaches steady-state after 125 hours, achieving a minimum temperature of 79 K. (the structure cool down from 8 points)

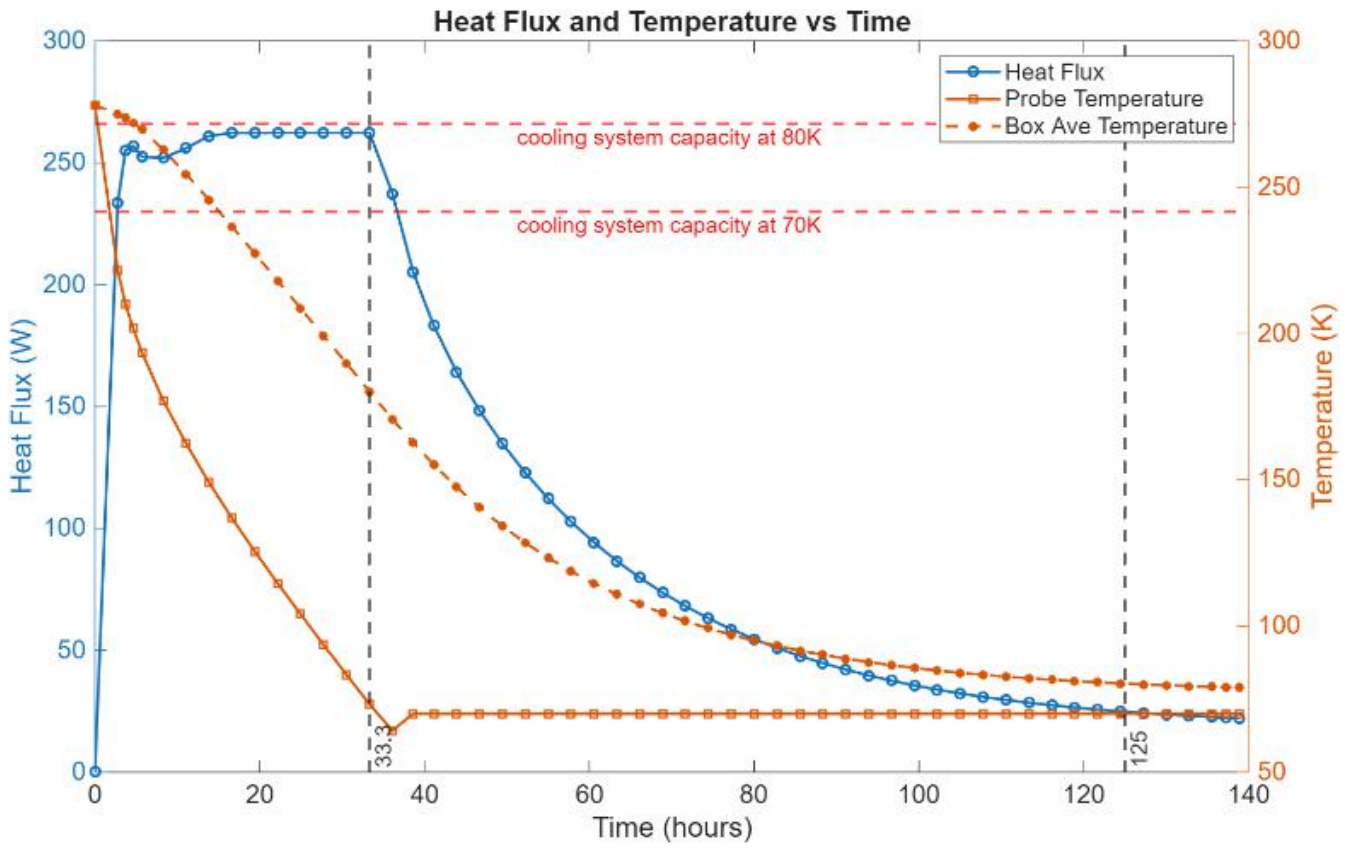

**Fig. 13** average heat flux extracted by each cooling probe and the corresponding contact point temperature versus time during the transient cool-down process (the structure cools down from 8 points). The system reaches 70 K after ~34 hours, after which the applied heat flux decreases significantly to maintain this temperature. Steady-state conditions are achieved after approximately 125 hours, with the applied heat flux remaining below the cryogenic system's maximum capacity.

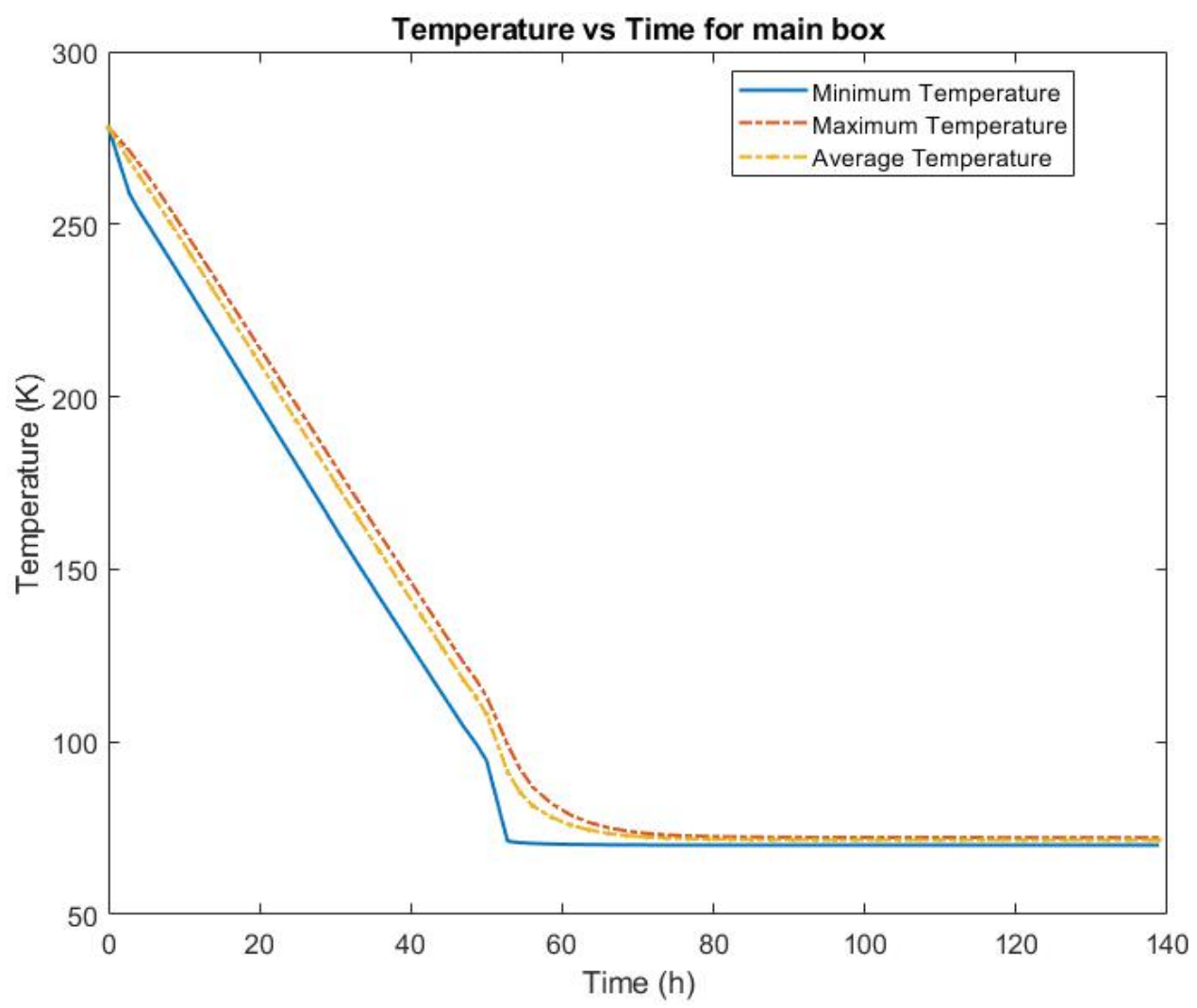

**Fig. 14** Average temperature of the main structure as a function of time during the cooling process. The system reaches steady-state after 65 hours, achieving a minimum temperature of 70 K. (The structure cools down from 16 points)

*2.Steady-state thermal analysis*

In the steady-state thermal analysis, the boundary conditions are like those used in the transient analysis, with the following differences. The structure is cooled exclusively by maintaining the probe contact points at 70 K. Internal heat powers are applied to represent the NEXUS and VESPER motorizations. Since the number of motors in NEXUS mode (as a worst-case scenario) is higher than in VESPER mode, only the NEXUS mode is considered for the steady-state analysis, with an assumed internal heat load of 10 W for the grism wheel and 1 W for the configurable slit system (CSS). All other boundary conditions remain consistent with the transient analysis.

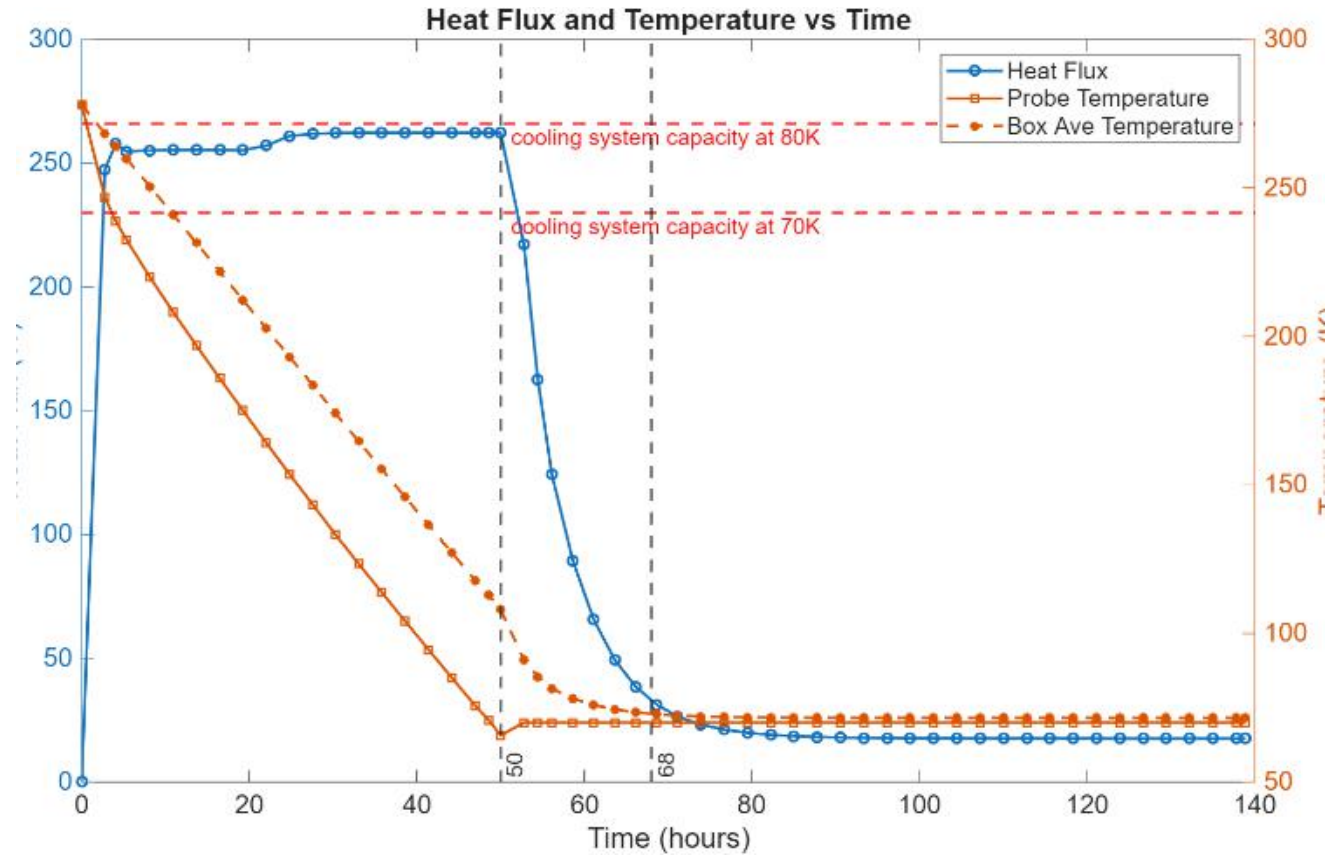

**Fig 15.** Average heat flux extracted by each cooling probe and the corresponding contact point temperature versus time during the transient cool-down process (the structure cools down from 16 points). The system reaches 70 K after ~55 hours, after which the applied heat flux decreases significantly to maintain this temperature. Steady-state conditions are achieved after approximately 68 hours, with the applied heat flux remaining below the cryogenic system's maximum capacity.

The temperature distribution obtained from the steady-state analysis is shown in Figure 16. The results indicate that a total cooling power of 157 W is required to maintain the structure in steady-state conditions.

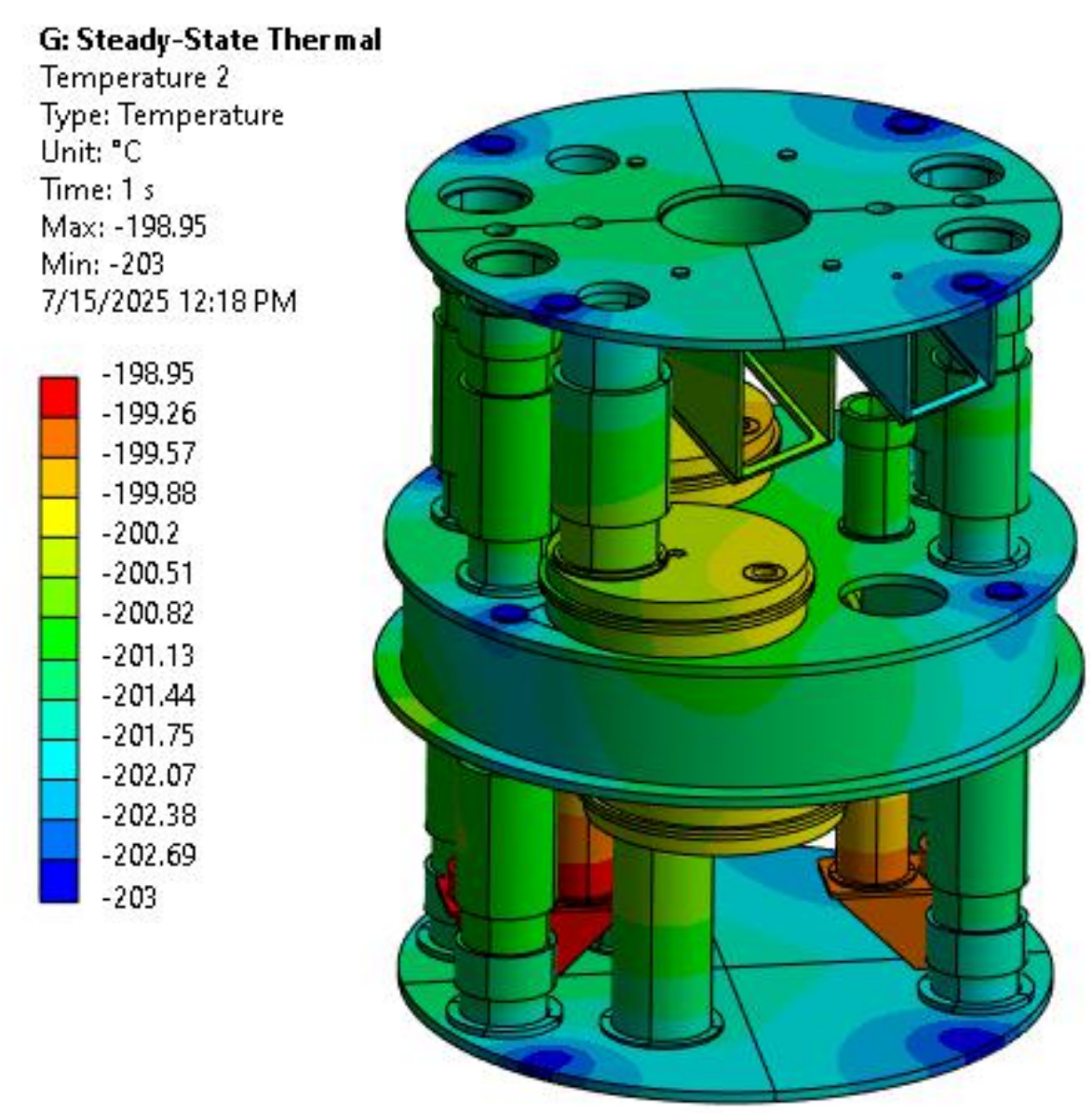

**Fig. 16** Steady-state temperature distribution of the structure under NEXUS mode.

### B. Structural analysis

The structural deformation induced by thermal stress must remain below the optical error budget to guarantee that SHARP operates without performance degradation. To verify structural and optical integrity, the deformations of the optical elements were evaluated under steady-state thermal conditions (Thermal distribution applied as load) combined with gravitational loading. The system is rigidly constrained on the middle steel ring of the external structure.

Figure 17 illustrates the overall deformation of the SHARP structure in the steady state analysis. As described earlier, the optical elements are modelled as point masses rigidly linked to the relative supporting structures; at the mounting interface, and therefore, the displacements of these mass points are interpreted as the deformations of the corresponding optical components. The results for the individual optical elements are summarised in Table 7for NEXUS optic elements and in Table 8 for VESPER optic elements (results for one channel are presented). The results of the structural analysis show that the deformation of the optical elements, particularly the most sensitive components, remains well within the allocated optical error budget. Consequently, a significant portion of this margin can be reallocated to accommodate manufacturing and alignment tolerances at room temperature. Therefore, SHARP can maintain its required accuracy using passive alignment to correct manufacturing and assembly errors, although the implementation of active compensators remains a possible option. The detailed alignment procedure will be studied and described in a dedicated paper.

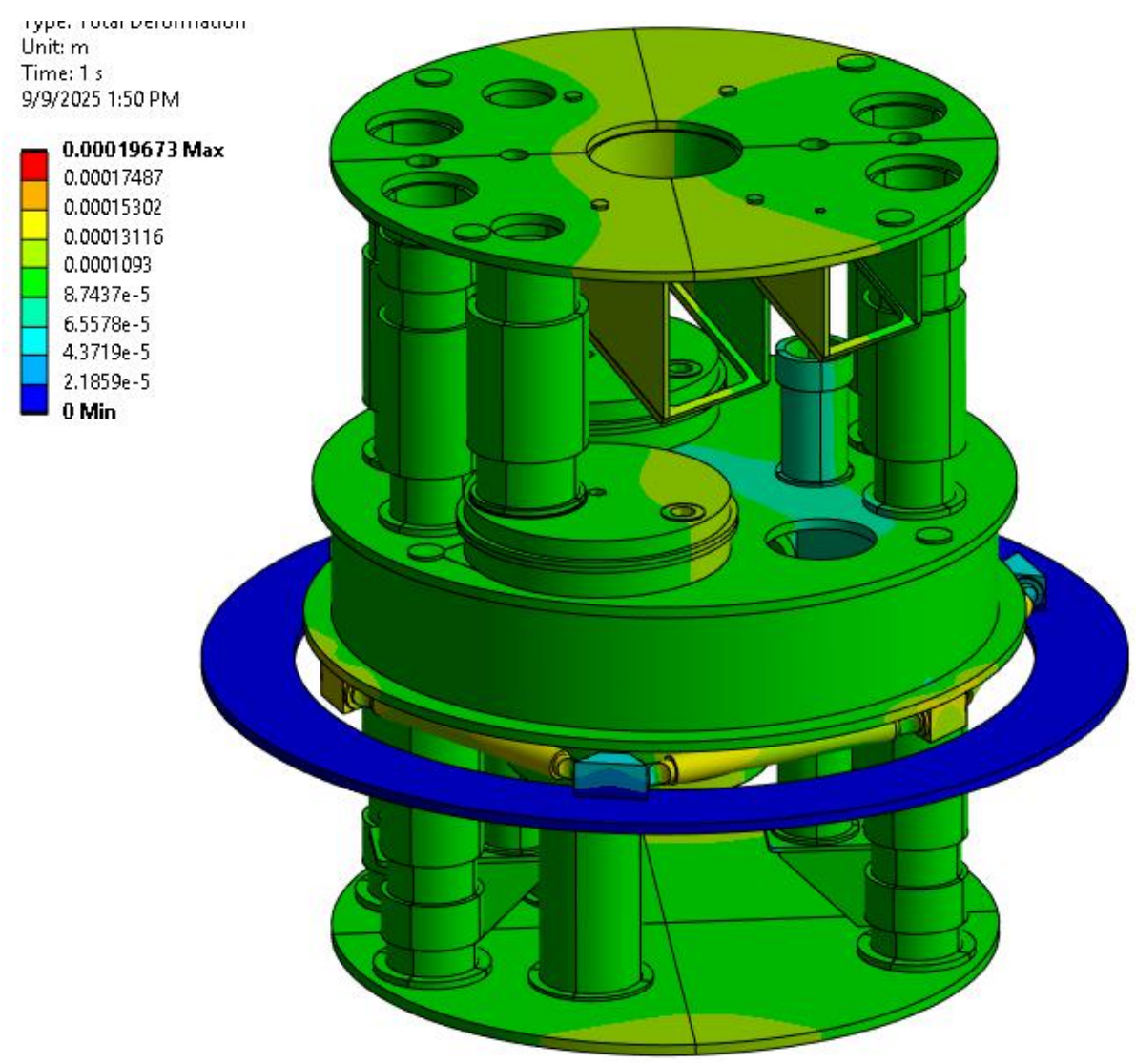

**Fig. 17** Steady-state structural deformation of the SHARP system under thermal loading. Optical elements are modelled as point masses, and their displacements are evaluated relative to the error budget.

**Table 7: Structural Deformation Results of NEXUS Optical Elements (Steady-State).**

| Optical Element | Decenter (mm) | Tilt (°) | Despace (mm) | Status (Pass/Fail) |
|---|---|---|---|---|
| Field Lens L1 | 0.0162 | 0.0006 | 0.0000 | |
| Mirror M1 | 0.0189 | 0.0007 | 0.0002 | Pass |
| Mirror M2 | 0.0243 | 0.0026 | 0.0048 | Pass |
| Mirror M3 | 0.0127 | 0.0008 | 0.0137 | Pass |
| Field Lens L2 | 0.0921 | 0.0003 | 0.0052 | Pass |
| Field Lens L3 | 0.0921 | 0.0003 | 0.0000 | Pass |
| Dichroic D1 | 0.0924 | 0.0003 | 0.0002 | Pass |
| Field Lens L4 | 0.0926 | 0.0004 | 0.0001 | Pass |
| Field Lens L5 | 0.0926 | 0.0004 | 0.0000 | Pass |
| Dichroic D2 | 0.0944 | 0.0005 | 0.0023 | Pass |
| Mirror M4 | 0.0964 | 0.0006 | 0.0001 | Pass |
| Field Lens L6 | 0.0960 | 0.0017 | 0.0109 | Pass |
| Field Lens L7 | 0.0960 | 0.0017 | 0.0000 | Pass |
| Dichroic D3 | 0.0978 | 0.0002 | 0.0095 | Pass |
| Mirror M5 | 0.0982 | 0.0001 | 0.0000 | Pass |
| C1-L1 | 0.0014 | 0.0014 | 0.0047 | Pass |
| C1-L2 | 0.0013 | 0.0014 | 0.0000 | Pass |
| C1-L3 | 0.0080 | 0.0012 | 0.0003 | Pass |
| C1-L4 | 0.0096 | 0.0012 | 0.0004 | Pass |
| C1-L5 | 0.0141 | 0.0013 | 0.0002 | Pass |
| C1-L6 | 0.0148 | 0.0013 | 0.0001 | Pass |
| C2-L1 | 0.0036 | 0.0008 | 0.0048 | Pass |
| C2-L2 | 0.0037 | 0.0009 | 0.0002 | Pass |
| C2-L3 | 0.0074 | 0.0007 | 0.0003 | Pass |
| C2-L4 | 0.0086 | 0.0006 | 0.0002 | Pass |
| C2-L5 | 0.0091 | 0.0006 | 0.0000 | Pass |
| C2-L6 | 0.0095 | 0.0006 | 0.0001 | Pass |
| C3-L1 | 0.0042 | 0.0002 | 0.0096 | Pass |
| C3-L2 | 0.0034 | 0.0002 | 0.0000 | Pass |
| C3-L3 | 0.0003 | 0.0002 | 0.0004 | Pass |
| C3-L4 | 0.0014 | 0.0003 | 0.0000 | Pass |
| C3-L5 | 0.0045 | 0.0003 | 0.0001 | Pass |
| C3-L6 | 0.0094 | 0.0002 | 0.0004 | Pass |
| C4-L1 | 0.0042 | 0.0014 | 0.0077 | Pass |
| C4-L2 | 0.0034 | 0.0014 | 0.0000 | Pass |
| C4-L3 | 0.0003 | 0.0013 | 0.0004 | Pass |
| C4-L4 | 0.0014 | 0.0013 | 0.0000 | Pass |
| C4-L5 | 0.0045 | 0.0012 | 0.0001 | Pass |
| C4-L6 | 0.0094 | 0.0012 | 0.0004 | Pass |

**Table 8: Structural Deformation Results of VESPER Optical Elements (Steady-State).**

| Optical Element | Decenter (mm) | Tilt (°) | Despace (mm) | Status (Pass/Fail) |
|---|---|---|---|---|
| **channel 1** | | | | |
| Mirror & Field Lens M1,2 L1,2 | 0.10002 | 0.00066 | 0.00864 | Pass |
| Mirror M3 | 0.10644 | 0.00104 | 0.00103 | Pass |
| Field Lens L3,4,5,6 | 0.00695 | 0.00042 | 0.00553 | Pass |
| Mirror M4 | 0.00774 | 0.00033 | 0.00081 | Pass |
| Mirror M5 | 0.00536 | 0.00027 | 0.00067 | Pass |
| Field Lens L7,8 | 0.00567 | 0.00043 | 0.00003 | Pass |
| Slicer bracket | 0.00415 | 0.00063 | 0.00192 | Pass |
| **Camera 1** | | | | Pass |
| Lens CL1 | 0.00212 | 0.00094 | 0.09939 | Pass |
| Lens CL2,3 | 0.00094 | 0.00103 | 0.00017 | Pass |
| GRISM | 0.00135 | 0.00104 | 0.00000 | Pass |
| Lens CL4 | 0.00295 | 0.00105 | 0.00006 | Pass |
| Lens CL5,6 | 0.00395 | 0.00106 | 0.00000 | Pass |
| Lens Cl7 | 0.00501 | 0.00107 | 0.00005 | Pass |
| Lens CL8 | 0.00872 | 0.00107 | 0.00005 | Pass |
| Lens CL9 | 0.01059 | 0.00107 | 0.00003 | Pass |
| **Camera 2** | | | | Pass |
| Lens CL1 | 0.00120 | 0.00110 | 0.09837 | Pass |
| Lens CL2,3 | 0.00245 | 0.00107 | 0.00032 | Pass |
| GRISM | 0.00383 | 0.00107 | 0.00002 | Pass |
| Lens CL4 | 0.00556 | 0.00107 | 0.00010 | Pass |
| Lens CL5,6 | 0.00656 | 0.00107 | 0.00001 | Pass |
| Lens Cl7 | 0.00760 | 0.00108 | 0.00010 | Pass |
| Lens CL8 | 0.01117 | 0.00108 | 0.00007 | Pass |
| Lens CL9 | 0.01300 | 0.00111 | 0.00003 | Pass |
| **Camera 7** | | | | Pass |
| Lens CL1 | 0.00603 | 0.00039 | 0.10045 | Pass |
| Lens CL2,3 | 0.00626 | 0.00019 | 0.00019 | Pass |
| GRISM | 0.00651 | 0.00017 | 0.00001 | Pass |
| Lens CL4 | 0.00672 | 0.00015 | 0.00008 | Pass |
| Lens CL5,6 | 0.00684 | 0.00014 | 0.00002 | Pass |
| Lens Cl7 | 0.00693 | 0.00010 | 0.00010 | Pass |
| Lens CL8 | 0.00708 | 0.00010 | 0.00005 | Pass |
| Lens CL9 | 0.00715 | 0.00011 | 0.00003 | Pass |
| **Camera 8** | | | | Pass |
| Lens CL1 | 0.01680 | 0.00156 | 0.10127 | Pass |
| Lens CL2,3 | 0.01401 | 0.00187 | 0.00065 | Pass |
| GRISM | 0.01147 | 0.00189 | 0.00001 | Pass |
| Lens CL4 | 0.00836 | 0.00190 | 0.00006 | Pass |
| Lens CL5,6 | 0.00667 | 0.00192 | 0.00002 | Pass |
| Lens Cl7 | 0.00505 | 0.00188 | 0.00001 | Pass |
| Lens CL8 | 0.00458 | 0.00189 | 0.00003 | Pass |
| Lens CL9 | 0.00718 | 0.00177 | 0.00010 | Pass |

## 3. Conclusion

In this work, a comprehensive structural, thermal, optical performance (STOP) analysis of the SHARP near-infrared spectrograph has been presented. Transient thermal simulations were carried out to characterize the cooldown process, estimate the required cooling power, and determine the near-optimal number and locations of the cooling probes and thermal bridges. Steady-state thermal analysis was subsequently performed to evaluate the thermal distribution and temperature gradients within the system, followed by structural analysis to assess the resulting stresses and deformations. Optical sensitivity and tolerance analyses were performed independently using Ansys Zemax and Monte Carlo simulations. The obtained optical tolerances were then compared with the thermo-mechanical deformations predicted by the structural and thermal analyses. The results confirmed that SHARP is capable of maintains optical performance within the specified error budgets when subjected to expected thermal and mechanical loads just using passive alignment. Furthermore, by comparing the structural deformations with the optical error budget, the allowable alignment-error margins at room-temperature assembly conditions could also be estimated. These findings confirm that SHARP could reliably achieve its stringent optical requirements under near-realistic operational conditions, providing confidence in both its opto-mechanical design and thermal stability, although it will need precise calibration and assemblies that could be the most challenging part of SHARP. More broadly, this work emphasises the importance of STOP methodologies in the development of next-generation cryogenic astronomical instruments, where multidisciplinary coupling between structural, thermal, and optical domains is essential to guarantee performance.

Future work will focus on experimental validation of the numerical models through prototype testing under cryogenic conditions. In addition, integration of SHARP with MCAO-assisted telescope systems will be studied to evaluate system-level performance under realistic observing scenarios.

## Acknowledgments

HM, PS and the SHARP team acknowledge the support from the INAF Techno-Grant 2022, SHARP - 1.05.12.02.01; and the INAF Large-Grant 2024, SHARP - 1.05.24.01.01.

## Disclosures

The authors declare no conflicts of interest.

## Data availability

All data in support of the findings of this paper are available within the article. More information can be found on this website: https://sharp.brera.inaf.it/.